\documentclass[%
 reprint,
superscriptaddress,
 amsmath,amssymb,
 aps,
]{revtex4-2}

\usepackage{graphicx}
\usepackage{dcolumn}
\usepackage{bm}
\usepackage{xcolor} 
\usepackage{soul} 
\usepackage[mathlines]{lineno}

\usepackage{mathtools} 

\begin{document}

\preprint{APS/123-QED}

\title{Quantum Thermometry with single molecules in portable nanoprobes}

\author{V. Esteso}
 \altaffiliation[Both authors contributed equally]{.}
\affiliation{%
National Institute of Optics (CNR-INO), c/o LENS via Nello Carrara 1, Sesto F.no 50019, Italy}
\affiliation{European Laboratory for Non-Linear Spectroscopy (LENS), Via Nello Carrara 1, Sesto F.no 50019, Italy}
\affiliation{Departamento de Física de la Materia Condensada, ICMSE-CSIC, Universidad de Sevilla, Apdo. 1065, 41080, Sevilla, Spain
}%

\author{R. Duquennoy}
 \altaffiliation[Both authors contributed equally]{.}
\affiliation{%
National Institute of Optics (CNR-INO), c/o LENS via Nello Carrara 1, Sesto F.no 50019, Italy}
\affiliation{%
Physics Department, University of Naples, via Cinthia 21, Fuorigrotta 80126, Italy}

\author{R. C. Ng}%
\affiliation{%
Catalan Institute of Nanoscience and Nanotechnology (ICN2), CSIC and BIST, Campus UAB, Bellaterra, 08193 Barcelona, Spain}

\author{M. Colautti}%
\affiliation{%
National Institute of Optics (CNR-INO), c/o LENS via Nello Carrara 1, Sesto F.no 50019, Italy}
\affiliation{European Laboratory for Non-Linear Spectroscopy (LENS), Via Nello Carrara 1, Sesto F.no 50019, Italy}
 
\author{P. Lombardi}%
\affiliation{%
National Institute of Optics (CNR-INO), c/o LENS via Nello Carrara 1, Sesto F.no 50019, Italy}
\affiliation{European Laboratory for Non-Linear Spectroscopy (LENS), Via Nello Carrara 1, Sesto F.no 50019, Italy}

\author{G. Arregui}%
\affiliation{%
Catalan Institute of Nanoscience and Nanotechnology (ICN2), CSIC and BIST, Campus UAB, Bellaterra, 08193 Barcelona, Spain}
\affiliation{DTU Electro, Department of Electrical and Photonics Engineering, Technical University of Denmark, Ørsteds Plads 343, Kgs. Lyngby, DK-2800, Denmark
}%

\author{E. Chavez-Angel}%
\affiliation{%
Catalan Institute of Nanoscience and Nanotechnology (ICN2), CSIC and BIST, Campus UAB, Bellaterra, 08193 Barcelona, Spain}
 
\author{C. M. Sotomayor Torres}%
\affiliation{%
Catalan Institute of Nanoscience and Nanotechnology (ICN2), CSIC and BIST, Campus UAB, Bellaterra, 08193 Barcelona, Spain}
\affiliation{ICREA, Passeig Lluis Companys 23, 08010 Barcelona, Spain
}%

\author{P. D. Garcia}%
\affiliation{%
Instituto de Ciencia de Materiales de Madrid, (ICMM) Calle Sor Juana Inés de la Cruz 3, 28049 Madrid, Spain. Consejo Superior de Investigaciones Científicas (CSIC)
}%
\author{M. Hilke}%
\affiliation{%
 Department of Physics, McGill University Montréal, QC, Canada H3A 2T8}
\affiliation{Department of Physics, University of Florence, Via Sansone 1 50019, Sesto Fiorentino, Italy
}%

\author{C. Toninelli}%
\email{Corresponding author: toninelli@lens.unifi.it}
\affiliation{%
National Institute of Optics (CNR-INO), c/o LENS via Nello Carrara 1, Sesto F.no 50019, Italy}
\affiliation{European Laboratory for Non-Linear Spectroscopy (LENS), Via Nello Carrara 1, Sesto F.no 50019, Italy}




\date{\today}

\begin{abstract}
Understanding heat transport is relevant to develop efficient strategies for thermal management in microelectronics for instance, as well as for fundamental science purposes.
However, measuring temperatures in nanostructured environments and in cryogenic conditions remains a challenging task, that requires both high sentitivity and a non-invasive approach. Here we present a portable nanothermometer based on a molecular two-level quantum system that operates in the 3 - 30 K temperature range, with excellent temperature and spatial resolutions on the order of mK and $\mu$m, respectively. We validate the performance of this molecular thermometer on nanostructures, by estimating the thermal conductivity of a patterned silicon membrane. In addition, we demonstrate the two-dimensional temperature mapping of a patterned surface via the simultaneous spectroscopy of all thermometers deposited on a sample. These results demonstrate the potential of this molecular thermometer to explore thermal properties and related phenomena at cryogenic temperatures.

\end{abstract}

\maketitle


\section{\label{sec:level1}Introduction}

The precise measurement of local temperature is crucial for efficient thermal management and control of heat conduction. While this control is a major goal at the macroscale \cite{PASUPATHY200839,YANG20211,zhang2022advanced,feng2022emerging}, it is also extremely relevant in microelectronics \cite{mathew2022review,li2022thermal,chen2022scalable,LAWAG2022105602}, biological environments \cite{brites2012thermometry,kim2015microscale}, and even in fundamental quantum science \cite{jevtic2015single,maire2017heat,du2019quantum,ccaugin2021thermal,feng2021thermal}. Despite its technological relevance, efficient thermometry methods are challenging, especially in more complex environments such as those in which cryogenic temperatures or nanoscale dimensions are involved. Interestingly, under these conditions a multitude of different non-Fourier heat conduction transport regimes are believed to exist \cite{chen2021non}, including ballistic, Casimir-Knudsen \cite{casimir1938note,knudsen1950kinetic} and phonon hydrodynamics \cite{cepellotti2015phonon}. An ideal thermal probe requires minimal invasitvity to avoid altering the measured properties, while simultaneously possessing sufficient sensitivity and spatial resolution, in order to map temperature fluctuations and defects with nanoscale resolution. It also requires portability of the sensor, such that the probe can be brought in proximity with the target system with minimal perturbation.

Among currently available nano- and micro- thermometric methods \cite{brites2012thermometry,kim2015microscale,quintanilla2018guiding}, the most prominent ones are those based on thermo-electric effects \cite{reverter2021tutorial,qiao2022extension} or on optical techniques \cite{liu2015mixed,bradac2020optical,chen2020optical,zhang2023two} such as the frequency shift measurement in Raman signals \cite{reparaz2014novel,chavez2014reduction}, or the measurement of  the reflectivity change in the case of time and frequency-resolved thermoreflectance techniques \cite{doi:10.1063/5.0020239,zen2014engineering}. The former approaches enable high sensitivity reaching down to the mK range. However, they tend to be invasive at the nanoscale and are lacking in their ability of providing information with high spatial or 2D resolution. Raman-based methods instead allow for fast and spatially resolved read-out but are inefficient as temperatures decrease below a certain temperature depending on the system (typically below 100 K). Moreover, they are only functional for materials with Raman-active modes. Finally, thermoreflectance techniques suffer from limited sample-flexibility and often require the use of a transducer layer interacting with the measured samples. \cite{jang2010thickness}. 


Interesting results have been matured in the context of nanoparticle-based temperature detection, including those relying on luminescence measurements \cite{chihara2019biological,dramicanin2020trends,anaya2015thermal}. Such classical approaches are though typically optimized for room temperature operation.

The detection of temperature-induced effects on quantum systems allows theoretically to simultaneously reach both low temperature ranges and high sensitivity \cite{mitchison2020situ}, providing an alternative and promising strategy to measure the temperature of a target and its surrounding. A notable example is that of temperature-induced decoherence on a two-level system (qubit), which leaves the thermal bath largely unperturbed, as no actual thermalization or energy exchange take place. Furthermore, quantum metrology strategies can be applied to identify optimal measurements, yielding the best precision and accuracy given the interaction Hamiltonian \cite{razavian2019quantum,xie2017optimal}. However, the extreme sensitivity of quantum probes to the environment complicates the decoupling of temperature effects from other sources of decoherence and hinders their deployment within portable solid state devices. 

In this paper we present a non-invasive nanothermometer based on quantum  emitters hosted in molecular crystals and employed as quantum probes. These fluorescent molecules are embedded as interstitial impurities in submicron-sized crystals, which are then deterministically positioned on the sample at the microscale, and investigated under cryogenic conditions. The precise estimation of the local temperature is obtained from the measurement of the molecules' optical transition linewidth.

The molecular thermometer is calibrated by comparing the response of several molecular emitters on different surfaces as a function of the cryostat temperature. By exploiting the absorption of a laser beam impinging on the sample at a controlled power and position, heat conduction in nanostructured silicon membranes is studied as an interesting testbed, and characterized with unprecedented sensitivity in the temperature range from $3$ to $30$ K. Temperature maps are also acquired with discrete sampling of the surface over approximately $20$x$20$ $\mu m^2$ (although the area could be as large as a few mm$^2$).


\section{\label{sec:level2}Concept and Experimental Setup}


\begin{figure*}
\includegraphics[width=1.5\columnwidth]{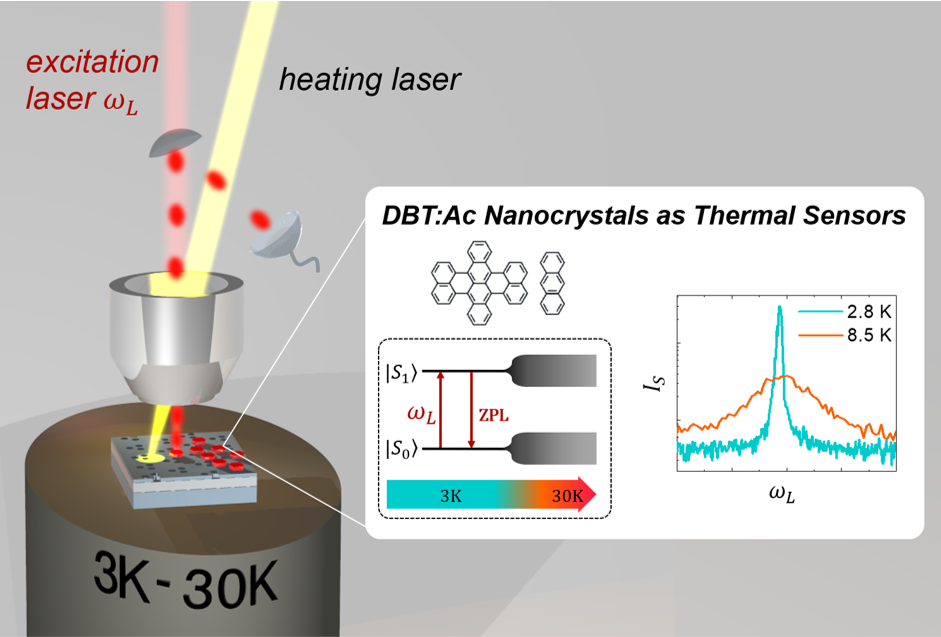}
\caption{\label{fig:1}Illustration of the experimental setup and working principle. From left to right: sketch of the experimental setup where an excitation laser (red beam) interrogates the molecular thermometers (red disks on the sample surface) consisting of dibenzoterrylene (DBT) molecules embedded in anthracene (Ac) nanocrystals deposited on the sample, while the heating laser (yellow beam) acts as a controllable heating source of the sample at a given distance from the DBT:Ac nanocrystal. The Stokes-shifted emission intensity (I$_S$) of the molecular thermometer is collected as a function of $\omega_L$ giving rise to broader Lorentzian peaks the higher the temperature at the sampled point. In the center, the Jablonski diagram of the energy levels of DBT molecules embedded in Ac nanocrystals is represented, where $|$S$_0 \rangle$ and $|$S$_1 \rangle$ signify the singlet ground and excited states, respectively. At low temperatures, the energy levels are well defined. At higher temperatures, the energy levels, and by consequence the transitions, broaden (see graph on the right). Calibration curves relating this broadening to temperature were previously measured by fixing the sample temperature in the cryostat from 3 to 32 K. With such information, discrete sampling with high spatial accuracy (a few hundreds of nanometers, which is the size of the nanocrystals) can be performed.}
\end{figure*}

Polyaromatic hydrocarbon chromophores embedded in appropriate host matrices and cooled down to cryogenic temperatures can be described in first approximation as two-level systems capable of emitting single photons with high brightness, purity, long-term photostability, and indistinguishability  \cite{toninelli2021single}. In particular, here we consider dibenzoterrylene (DBT) molecules embedded in anthracene (Ac) nanocrystals \cite{pazzagli2018self,nicolet2007single} as molecular thermometers. Fig. 1 shows the Jablonski diagram for the energy levels of DBT:Ac nanocrystals, with $S_0$ and $S_1$ being the ground and the excited electronic singlet states, respectively. The zero-phonon line (ZPL) is around 785 nm and its linewidth is lifetime-limited at low temperatures with a full width high maximum $FWHM = \Gamma_1/2\pi = 1/(2\pi\tau_1)$  $\simeq 50$ MHz, with $\tau_1$ being the radiative lifetime \cite{pazzagli2018self}. In addition to a purely electronic transition, the molecular excited state can decay via internal vibrational states, yielding fluorescence at longer wavelengths. Moreover, electron-phonon coupling with mechanical excitations of the matrix gives rise to the so-called phonon side bands in the emission spectrum as a first order effect, and induces decoherence of the molecular transition dipole moment as a second order effect \cite{hochstrasser1972phonon,norambuena2016microscopic,reitz2020molecule,clear2020phonon}. The molecule thereby can also be interpreted as acute sensors of its local environment, which, at a given temperature, is characterized by the phonon spectral density of states and the relative average occupation number, as given by the Boltzamnn distribution. The exact optomechanical interactions that occur between photons, localized mechanical excitation of the molecules, and phonons is a rich and active area of research \cite{chen2021continuous,gurlek2021engineering}. Here, we focus more specifically on exploiting the temperature-dependent optical response that results from these interactions. In particular, we will explore the molecular dynamics by considering that the ZPL transition linewidth (FWHM) is expected to follow a close-to-exponential increase with temperature, according to \cite{clear2020phonon} (additional details are provided below).


Towards this goal, excitation spectroscopy is used, scanning the laser frequency $\omega_L$ across the ZPL transition and collecting the red-shifted fluorescence $I_S$ after a long-pass filter. The experiments are done using a confocal microscope in an epifluorescence configuration, as sketched on the left in Fig.\ref{fig:1}, (more details on the optical setup are found in Refs. \cite{pazzagli2018self,duquennoy2022real}). The setup allows for detection with either an electron multiplied charge coupled device (EMCCD) camera or with single photon avalanche photodiodes (SPADs). Two SPADs are arranged in the Hanbury-Brown Twiss setup to detect the single molecule intensity autocorrelation function (see Fig. S1 in the  Supplemental Material).

The characteristic FWHM of the Lorentzian profile obtained from excitation spectroscopy is estimated under different testing conditions. In a first experiment, the temperature of the cyostat (a Montana closed-cycle cryostation) is controlled by local heaters and measured with a thermo-couple placed on the cold finger. This setup is used for calibration, as discussed further below.


In a second series of experiments we use an extra laser beam with a central wavelength of $\lambda= 767$ nm focused on the sample as a local heat source (heating laser in Fig.\ref{fig:1}). We set up two different configurations, referred to as type I and II from here onwards. In type I configuration (see Sec. \ref{sec:level3_2}), the heating laser position can be displaced with respect to a DBT:Ac nanocrystal probe (molecular thermometer), while single molecule excitation spectroscopy is performed on said probe. From the linewidth broadening, the local temperature on the surface is estimated as a function of the distance from the heating source. Following this approach, thermal properties (including the heat conduction) of complex materials can be studied at low temperature and, thanks to the nanocrystals' submicron size, also with high spatial resolution. Alternatively, in type II configuration (see Sec. \ref{sec:level3_3}), the heating laser remains static at a specific position on the surface. A large number of DBT:Ac nanocrystals distributed at random distances from the fixed heating source (see Methods section for more details on the deposition technique) in combination with wide-field excitation allows for spatial mapping of the temperature distribution on the surface. Indeed, a fluorescence map of all probed nanocrystals can be imaged on the EMCCD camera as a function of the scanning frequency of the excitation laser, yielding a spatial map of the ZPL of each molecular thermomenter. Such spatial information opens the possibility to study heat propagation and to characterize exotic samples at low temperatures, such as the case of complex materials with nanopatterned surfaces. 

\section{\label{sec:level3}Results and discussion}

\begin{figure*}
\includegraphics[width=1.5\columnwidth]{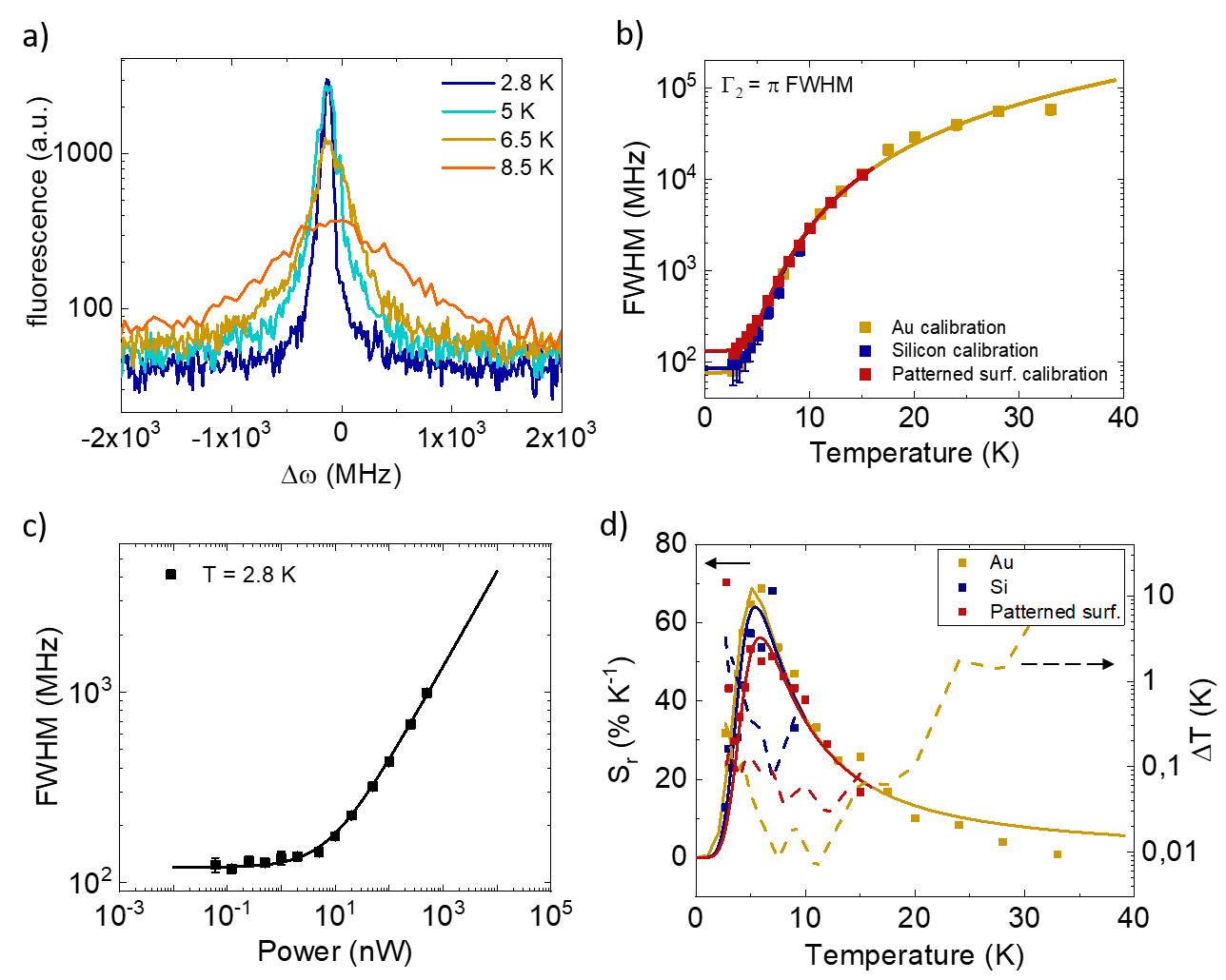}
\caption{\label{fig:2}Performance of the DBT:Ac nanocyrstal as a thermometer. a) Intensity of the Stokes-shifted emission of the DBT:Ac nanocrystal as a function of the frequency detuning ($\Delta\omega$) between the excitation laser and the ZPL of the molecule, at different sample temperatures as set by the cryostat for a silicon substrate. b) Calibration curves, i.e. linewidth vs sample temperature set by the cryostat, on gold (yellow), silicon (blue), and patterned silicon (red) surfaces. Squares show the experimental measurements with error bars and solid lines display the best fits to Eq. \ref{equation2}. c) Power dependence of the linewidth at $T$ = 2.8 K. d) Thermometer relative sensitivity (S$_r$) and temperature resolution ($\Delta$T) as a function of temperature. Squares and solids lines (color consistent with panel b), refer to the left axis while dashed lines refer to the right axis.}
\end{figure*}

\subsection{\label{sec:level3_1} Characterization of low-temperature single-molecule thermometers}

A typical example of the thermometer calibration is reported in Fig.\ref{fig:2}a), which displays the red-shifted fluorescence intensity from a single DBT molecule on a flat silicon surface as a function of the laser detuning with respect to the ZPL ($\Delta\omega = \omega_{ZPL}-\omega_L$). The molecule is excited confocally below saturation with a laser power, measured at the back entrance of the objective, $P=[0.3-5]$ nW as the temperature increases from $T$ = [2.8-11] K, and $P$ up to 100 nW for the highest temperatures. A clear broadening of the linewidth with increasing cryostat temperature is observed. The measurements are repeated for different molecules and sample surfaces to extract the specific associated calibration curves (FWHM vs temperature). The results for gold, silicon, and a patterned silicon membrane are reported in Fig.\ref{fig:2}b), covering the temperature range from $2.8$ to approximately $32$ $K$. The upper limit of the range of temperatures stems from the weak signal-to-noise ratio in fitting a Lorenztian peak at high temperatures. Additional examples of calibration curves are  provided as supplemental material. The experimental data (squares) and error bars are obtained as the average value and standard deviation from the average, respectively, from the Lorentzian fit of the molecule fluorescence as a function of detuning over at least four repeated laser scans. 
The best fit to the data is plotted in solid lines with the expression from Ref. \cite{clear2020phonon}

\begin{equation}
\Gamma_2 (T)= \Gamma_1/2+ \Gamma_2^* (T)
\label{equation1}
\end{equation}
where $\Gamma_2$ = $\pi\cdot$FWHM, $\Gamma_1$$= 1/\tau_1$ which is largely temperature independent, and $\Gamma_2^* (T)$ is the phonon-induced pure dephasing contribution. Clear et al. \cite{clear2020phonon} found that the latter can be expressed as
\begin{equation}
\begin{multlined}
\Gamma_2^* (T) = \mu \int_0^{\infty} d\omega \omega^6 n(\omega)(n(\omega)+1)    \cdot \\ \cdot
\int_0^{\pi} d\theta sin(\theta)(1+cos(\theta))^4 e^{-2\omega^2(1+cos(\theta))/\xi^2}
\label{equation2}
\end{multlined}
\end{equation}
with $n(\omega)$ being the Boltzmann phonon distribution function for a given $\omega$ frequency, and $\theta$ the angle between two phonons wavevectors. 
The above expression has been specifically developed to analytically deal with the electron-phonon interaction in the context of molecular crystals. The two fitting parameters, $\mu$  and $\xi$, which correspond to the amplitude and to the phonon cut off frequency, respectively, yield similar values for all three surfaces and different molecules ($\mu$ = ( $7.1 \pm 0.5 $) x $ 10^{-6}$ ps$^5$, and $\xi$ = ($6.1 \pm 0.3)$ ps$^{-1}$), leaving only $\Gamma_1$ as variable fitting parameter of $\Gamma_2(T)$. This means that estimating the $\mu$ and $\xi$ fitting parameters once for DBT:Ac crystals is sufficient for all cases, which is a huge simplification. Consequently, the measurement of $\Gamma_1$ for each DBT molecule that is used as a sensor is sufficient for calibration, without the need for consideration of the specific sample surface that is being analyzed.

In order to reach higher temperatures, we use gold substrates as the net gain for both excitation and collection efficiency allows for a better signal-to-noise ratio \cite{checcucci2017beaming} hence enabling a proper linewidth estimation even when the reduction of the absorption cross-section and the broadening of the line make it more difficult. More specifically, pump power is increased with temperature to maintain a constant excitation efficiency, corresponding to a saturation parameter below 1, for all the considered temperatures. The characteristic saturation behaviour of the system is captured in Fig. \ref{fig:2}c), showing the linewidth as a function of the excitation power in confocal excitation at 2.8 K. The data are fitted to $\Gamma_2(P) = \Gamma_2 \sqrt{1+\frac{P}{P_S}}$, which yields a typical saturation power $P_S=$ (7.8 $\pm$ 0.6) nW. All measurements presented in this work on silicon interfaces employ laser powers $\leq$ $5$ nW for the confocal configuration. These low power values ensure that the emission linewidth is not power broadened, representing a record level of low operating power, and consequently low invasivity for temperature measurements based on an optical technique. 

We now turn to  the main figures of merit for thermometers: the sensitivity and temperature resolution \cite{dramicanin2016sensing,quintanilla2018guiding}. A relative sensitivity is defined as $S_r=\frac{1}{\gamma}  \frac{d\gamma}{dT}\cdot 100$, expressed in $\% K^{-1}$ where $\gamma$ = FWHM, which is used as the indirect measurement of temperature. Fig. \ref{fig:2}d) summarizes the experimental (squares) and calculated sensitivities using Eq. 2 (solid lines) $S_r$ for three different surfaces. Regardless of the measured surface, the maximum $S_r$ is around $5$ to $7$ $K$ and reaches up to $\simeq70\%$ $K^{-1}$, among the highest $S_r$ reported for all-optical thermometers \cite{brites2012thermometry,zhou2016nanothermometry}. Dashed lines in Fig. 2d) correspond to the temperature resolution $\Delta$$T$,  given by the expression: $\Delta T$ =  $\frac{\Delta\gamma}{d\gamma/dT}$, where $\Delta\gamma$ is the experimentally estimated error bar on the linewidth at a given temperature (see above). Consequently, the temperature resolution is a measure of the minimum change in temperature that this technique can distinguish. Depending on the substrate, this amount varies from $0.01$ to $0.1$ $K$ over the temperature range $5 - 10 K$, being among the best temperature resolution reported at nanoscale \cite{brites2012thermometry}, as the collection efficiency and thus the error bars of the FWHM fitted values ($\Delta\gamma$) varies from substrate to substrate. Furthermore, the spatial resolution is given by the size of the nanocrystals, which is on average only a few hundreds of nanometers (see Ref. \cite{pazzagli2018self}), providing fine-grained access to local values of temperature and temperature gradients.

The overall performances of the molecular thermometer make it an excellent candidate for measuring heat transport under relevant environmental conditions, particularly those in which non-Fourier diffusion of heat is predicted to occur. This phenomenon tends to be difficult to investigate due to the characteristic low temperature range and the fragility of samples \cite{chen2021non}.

\begin{figure*}
\includegraphics[width=2\columnwidth]{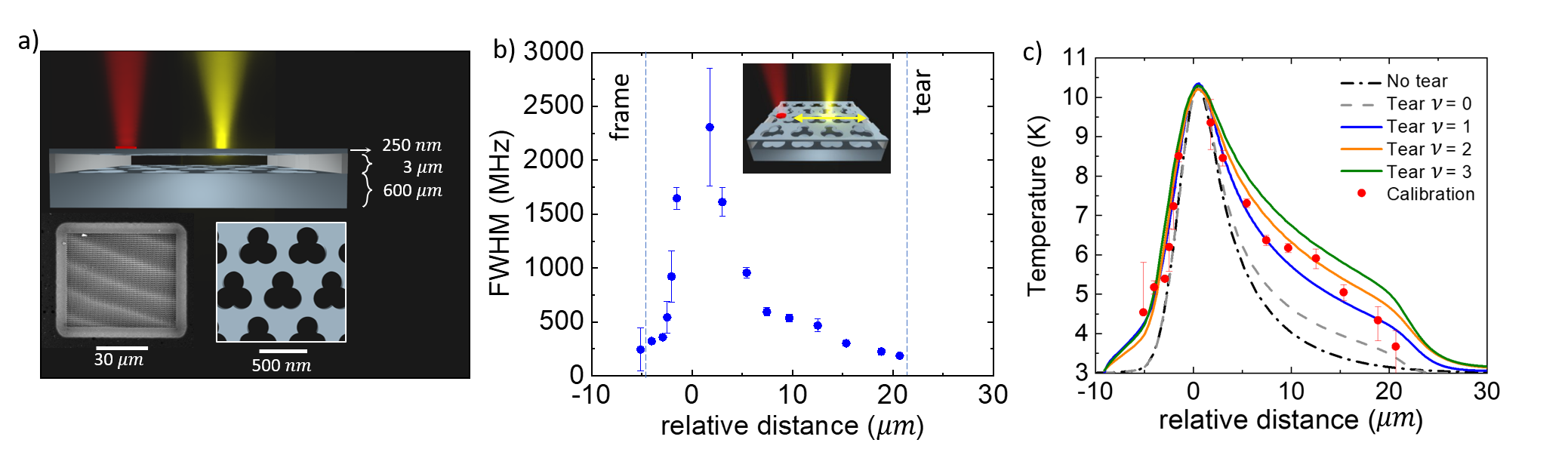}
\caption{\label{fig:3} Temperature profile on a silicon patterned membrane. a) Cross-section of the sample indicating the thickness of the membrane, the substrate and the space between them, together with a representative placement of the position of the lasers (red and yellow beams). A scanning electron microscope image of the complete membrane reveals the frame surrounding the membrane in the inset on the left. The inset on the right shows the shamrock-shaped pattern of the membrane. b) Linewidth vs the relative distance between the moving heating laser and the DBT:Ac nanocrystal. Error bars correspond to repeated laser scans as explained in Sec. \ref{sec:level3_1}. Inset shows a schematic of the configuration in which measurements are taken, interrogating only one nanocrystal while displacing the heat source. In addition, the dashed vertical lines indicate the location of the frame and the tear with respect to the position of the nanocrystal. c) Temperature profile as a function of the relative distance between the moving heating laser and the DBT:Ac nanocrystal. Red circles stand for temperatures estimated from experimental measurements in panel b) considering a calibration curve as described in \ref{sec:level3_1}. Solid lines represent simulated temperature profiles assuming different power laws of dependence of thermal conductivity on temperature, as well as the presence or lack thereof of the tear, as described in the text.
}
\end{figure*}

\begin{figure*}[t]
\includegraphics[width=1.5
\columnwidth]{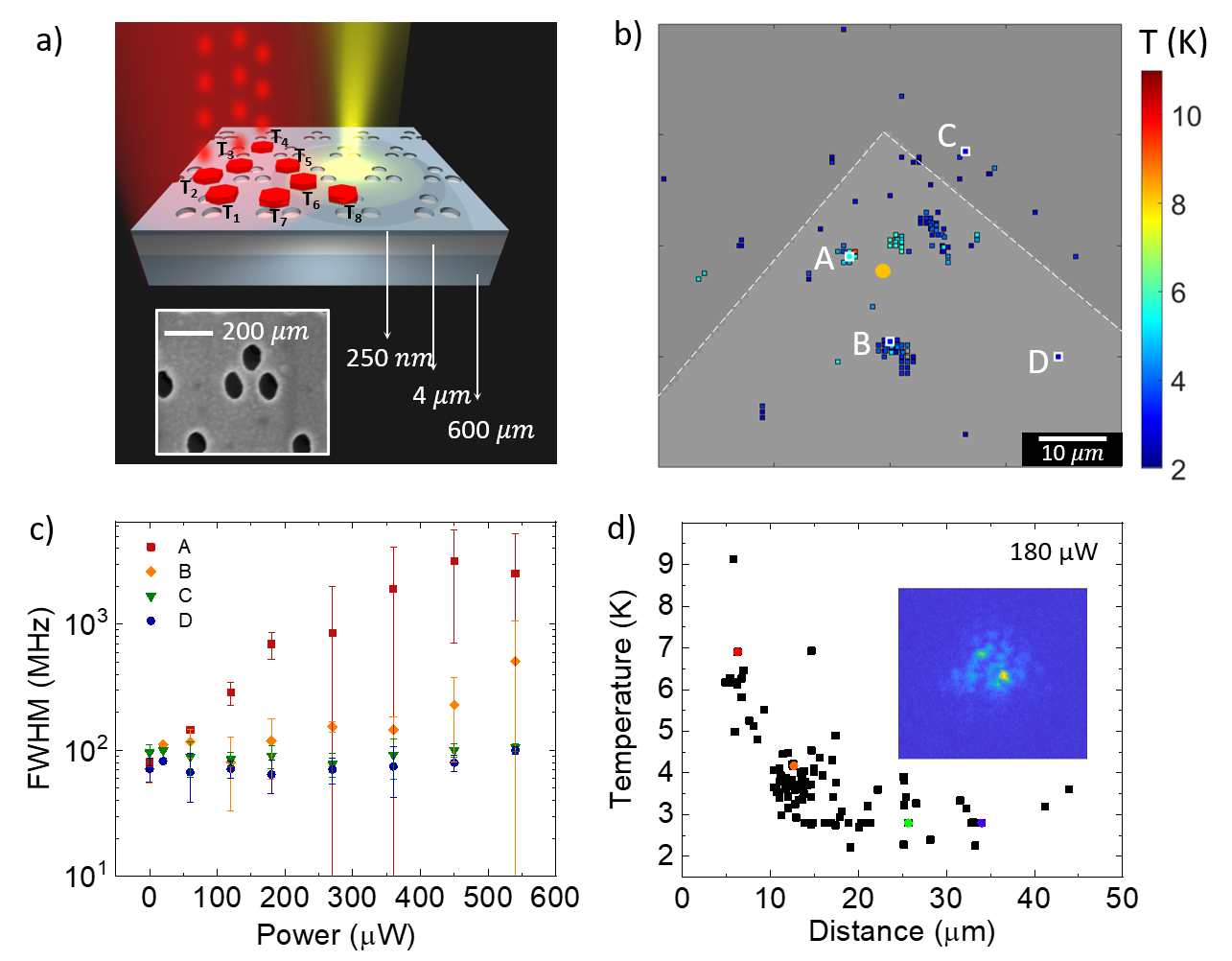}
\caption{\label{fig:4} 2D temperature mapping on a patterned silicon surface. a) Schematic of the measurement configuration in which the heating laser (yellow beam) remains static while the excitation laser (expanded red beam) allows collection of the photoluminescence (red spots of light) of several molecular thermometers (red hexagonal crystals on the sample) simultaneously, providing a temperature map of the sample. The sample consists of a SOI substrate with a 250 nm thick silicon device layer with the pattern shown in the inset, a 3 $\mu$m silicon dioxide buried oxide middle layer, and a 600  $\mu$m silicon substrate. b) Top view image of a corner of the patterned silicon surface. Heating laser and DBT:Ac nanocrystals are represented by a yellow circle and small squares, respectively. Color scale denotes the temperature measured by each molecular thermometer on the sample surface. c) Power dependence of the linewidth for four DBT:Ac nanocrystals labeled as A, B, C, and D, highlighted in white in panel b). d) Temperature as a function of the DBT:Ac nanocrystal distance with respect to the heating source at a given heating power of 180 $\mu$W. Colored squares correspond to labeled nanocrystals in panel c) with the same color code. Inset shows the speckle pattern of the heating laser on the sample surface.}
\end{figure*}

\subsection{\label{sec:level3_2} Measurements of heat propagation in nanopatterned silicon membranes}



As a representative example of a more complex surface, we study the heat propagation in a phononic crystal. The device is realized as a nanopatterned suspended silicon membrane with shamrock-shaped holes arranged in a periodic 2D triangular lattice. Such kinds of patterns have commonly been used to control vibrations in solids and to prevent the propagation of specific phononic frequencies which can be unwanted sources of noise and dephasing in quantum systems \cite{martinez1995sound,gorishnyy2005hypersonic,florez2022engineering}. Fig. 3a) displays a sketch of the sample's geometry with the corresponding thicknesses of the patterned membrane ($250$ nm), the silicon substrate ($600$ $\mu$m), and the space between the suspended membrane and the substrate (nominally 3 $\mu$m). The cross-sectional view also shows a generic position of both lasers. Specifically, we consider a suspended silicon membrane of size $50$ $\mu$m x $40$ $\mu$m with a non-patterned suspended silicon frame of $4.5$ $\mu$m of width (see inset on the left). The shamrock-shaped pattern with a periodicity of $500$ nm and a fill fraction of approximately $50$$\%$ is depicted in the inset on the right (see details on geometry in \cite{florez2022engineering} and more details of fabrication in the Methods Section).

The inset in Fig. 3b) shows the measurement configuration, as described in Section \ref{sec:level1}: the heating laser (yellow light) moves along a line containing the selected molecular thermometer (red hexagonal crystal), whereas the resonant laser (red light) is scanned in frequency and the photoluminescence is collected for each relative distance, to estimate the corresponding ZPL linewidth. The obtained linewidths are plotted in Fig. \ref{fig:3}b) as a function of the relative distance between the DBT molecule, i.e. the molecular thermometer, and the heating laser position. The linewidth broadens as the heating laser comes closer to the DBT molecule.  It is worth noting that the DBT molecule is only $5$ $\mu m$ away from the frame (which can be considered as thermal sink) and also the fact that in this specific case, the membrane ruptured during the nanocrystal deposition, resulting in a tear at 20 $\mu m$ away from the molecular thermometer (both distances are highlighted in Fig. \ref{fig:3}b) with a vertical dashed blue line and respective labels). These two factors are what give rise to the asymmetric shape of the linewidth profile. 

To interpret the experimental results, we numerically time integrate the heat equation iteratively, by updating the spatial distribution of the temperature and local thermal conductivity at every time step until we reach thermal equilibrium. This allows the heat equation to be solved by considering a non-constant thermal conductivity. The temperature dependence of the thermal conductivity is of the form: $\kappa \sim T^{\nu}$ with $\nu$ the exponent of the power law with temperature. The exact geometry is obtained from high resolution SEM images, which are then digitized for the simulation. The simulation uses a mesh size resolution of $77$ nm. A fixed bath temperature boundary condition of 3 K is considered outside the membrane's frame. In addition, an extra boundary condition is considered for the tear, which considers a vanishing heat flux at the tear. The results of the numerical simulation are then compared to the experimental data of the temperature as a function of the heat source position in Fig.\ref{fig:3}c). Red circles correspond to the temperature values estimated from the experimental points in Fig. \ref{fig:3}b), assuming the calibration curve described in Sec. \ref{sec:level3_1} with a lifetime-limited linewidth $\Gamma_1/2\pi$ $ = 196$ MHz, which was obtained with the heating laser off. Lines show the simulated temperature profile assuming the power law dependence of the heat conductance with temperature.  Dashed gray and dashed dot black lines display the theoretical profile with constant thermal conductivity 0.005 W/(m·K). Additionally, the former includes the boundary condition due to the tear, whereas the latter assumes there is not tear, i.e., the membrane is not ruptured. All other $T^{\nu}$ curves take into account the presence of a tear.

We find that the power law that best fits our experimental measurements corresponds to an exponent $\nu$ = $1$. The estimation of the sample thermal conductivity at $4$ K for $\nu$ = $1$ and $\nu$ = $2$ is ($0.0032$ $\pm$ $0.0006$) W/(m$\cdot$K) and $0.0015$ $\pm$ $0.0003$ W/(m$\cdot$ K), respectively, estimating a $1$$\%$ absorption of the impinging power. This percentage has been determined using the transfer matrix method in the multilayer structure shown in Fig. 3a), employing the optical properties of silicon at room temperature from Ref. \cite{schinke2015uncertainty}. Our estimation of $\kappa$ is in accordance with values in literature for nanostructured silicon membranes at 4 K \cite{nomura2022review}. For instance, reported values of thermal conductivity for silicon fishbone nanowires with 200 nm of periodicity and lengths ranging from 1 to 20 $\mu$m are $\kappa$ = [0.007-0.015] W/(m$\cdot$K)\cite{maire2015thermal}; whereas thermal conductivity for 145-nm thick phononic crystals based on disordered holes of diameter 170 nm, arranged in squared lattice with periodicity 300 nm are in the range $\kappa$ = [0.012-0.016] \cite{maire2015thermal}. In particular, the thermal conductivity for the most similar structure to ours, which consist in a 145-nm thick phononic crystal based on holes with periodicity around 250 nm and neck-size, i.e. the smallest distance between two voids, equal to 50 nm (which roughly corresponds with our geometry), is in the range of 0.005 to 0.010 W/(m$\cdot$K), for different geometrical arrangements \cite{anufriev2016reduction,anufriev2017heat}. Such results, which are of the same order of magnitude as our estimations, are smaller compared to the $\kappa$ value for an 145-nm thick unpatterned membrane: $\kappa$ = 0.050 W/(m$\cdot$K) \cite{nomura2015thermal}, and much smaller than silicon bulk at 4 K: $\kappa > 10^2$  W/(m$\cdot$K) \cite{glassbrenner1964thermal}. These results validate the use of such molecular two-level system as thermometer with high sensitivity and temperature resolution, via the direct measurement of the temperature dependent transitional linewidth broadening.







\subsection{\label{sec:level3_3} Two-dimensional temperature mapping at 3 K with multiple quantum thermometers.}

The ability to simultaneously measure temperature at different positions on a surface would allow the heating and dissipation processes of surfaces to be studied as a whole, rather than locally, also significantly reducing acquisition time. Towards this goal, the type II experimental configuration (see Fig. 4a)) is incorporated, with a large number of doped Ac crystals deposited onto a patterned silicon surface via drop-casting. The inset shows a scanning electron microscope image of the nanostructured silicon on insulator (SOI) surface, which is characterized by circular holes $45$ nm in radius, $250$ nm deep with a $500$ nm periodicty and a fill fraction of approximately of 15$\%$ (scale bar is $200$ nm). Fig. \ref{fig:4}b) illustrated the nanocrystal positions with the related temperature estimation as obtained using localization techniques and excitation spectroscopy performed with the EMCCD camera (as a guide for the eye, the corner of the structured area on the sample is denoted with white dashed lines). The laser heat source is then kept at a fixed position, marked with a yellow dot, which stands at different distances from the diverse crystals. This clearly shows the possibility to map temperature profiles with discrete sampling and high spatial accuracy in the range around $5$ K.

As representative cases, we select four DBT:Ac nanocrystals labeled A, B, C, and D for further analysis. These are $6.3$, $12.7$, $25.7$, and $34.0$ $\mu m$ away from the heating source, respectively. Fig. \ref{fig:4}c) displays how the linewidth of these molecules behave differently as a function of the heating laser power. In particular, the closer the DBT:Ac is to the heat source, the stronger the molecule ZPL linewidth scales with power. As described in Sec. \ref{sec:level3_1}, measuring $\Gamma_1$=1/$\tau_1$ of each nanocrystals with a unique frequency scan of the resonant laser in wide-field illumination and imaging the sample on the camera, determines the calibration curve for each DBT:Ac nanocrystal given the parameters $\mu$ and $\xi$ in equation (2). Fig. \ref{fig:4}d) presents the temperature estimated using tens of DBT:Ac nanocrystals as a function of their distance from the heat source, operated with a laser power $P_H=180$ $\mu W$. Nanocrystals A, B, C, D are highlighted with their corresponding color in Fig. \ref{fig:4}c). The large amount of temperature sensors on the surface allows for a reconstruction of a clear temperature profile showing the characteristic length scale of thermal transport on this surface. In this regard, the characteristic length scale is lower compared to Fig. 3c) as the structure is not suspended. The scattering of the data should be ascribed to the speckle pattern produced by the heating laser on the surface (see inset in Fig. 4d)), which increases spatial fluctuations of the temperature.

During the preparation of this manuscript we became aware of related work in Ref. \cite{chen2022ultralow}, where the authors used NV centers in diamond crystals as thermal probes.

\section{Conclusions}
We presented a single-molecule portable thermometer which allows for the measurement of temperature via the broadening of the zero-phonon molecule transition over the $3$ to $30$ K temperature range. Our molecular thermometer, dibenzoterrylene molecules embedded in an anthracene nanocrystal, shows a relative sensitivity of 70 $\%$ K$^{-1}$ around 5 K, a thermal resolution ranging from 0.1 to 0.01 K, and a spatial resolution of a few hundreds of nanometers, which is the size of the nanocrystals. Furthermore, it operates at low power, less than 5 nW in confocal configuration, making it one of the most outstanding molecular thermometers in terms of performance in this temperature range. We demonstrate the potential of the presented molecular thermometer by performing 2D temperature mapping of a nanostructured surface, and validate the results by determining the heat conduction in a patterned membrane by comparing experimental and simulated temperature profiles. These results pave the way for studying the thermal properties of materials beyond the Fourier diffusion theory of phonon propagation, due to the operating temperature range of a few Kelvin, the possibility of 2D temperature maps, and the feasibility of working on nanostructured samples.

\begin{acknowledgments}
 We would like to thank Prof. Michel Orrit and Robert Smit for providing us the dibenzoterrylene molecules.
 This work is funded by the EC under the FET-OPEN-RIA project STORMYTUNE (G.A. 899587), from the EMPIR pro- gramme (project 20FUN05, SEQUME), co-financed by the Participating States and from the European Union’s Horizon 2020 research and innovation program. Also financial support has been received from: PNRR MUR project PE0000023-NQSTI. V.E. acknowledges funding from European Union (NextGenerationEU), the Ministerio de Universidades of Spain, and the University of Seville under the Grant Margarita Salas. R.C.N. acknowledges funding from the EU-H2020 research and innovation program under the Marie Sklodowska Curie Individual Fellowship (Grant No. 897148). ICN2 is supported by the Severo Ochoa program, the Spanish Research Agency (AEI, grant no. SEV-2017-0706) and the CERCA Programme/Generalitat de Catalunya.
C. T. conceived the research. V.E., R.D. performed the experiments; M. H. developed the theoretical model and performed data analysis together with V.E.. The samples were prepared by R.N., G.A., under the supervision of P.G. and C.S.T. P. L. helped with the optical setup and E. C. A. discussed on the theory of thermal conduction. V.E. and C.T. wrote the paper with critical feedback from all authors. Figures were prepared by M.C.
\end{acknowledgments}

\appendix

\section{Methods}

\textbf{DBT:Ac nanocrystals preparation and deposition} 
DBT:Ac nanocrystals are formed in aqueous suspension. 50 $\mu$L of a 1:10$^7$ mixture of 1mM DBT in toluene and 5 mM Ac in acetone solution is injected into 2 ml of milli-Q water and sonicated for 30 min. Solvents and Ac were purchased from Sigma-Aldrich, water was deionized by a Mili-Q Advantage A10 system (18.2 m$\Omega$ cm at 25 deg), and DBT ($>$90$\%$ pure) was obtained by custom synthesis performed by the Dutch company Mercachem, nowadays called Symeres.
The resulting nanocrystals were then deposited on the surface of interest employing two deposition methods: micro-infiltration and drop-casting, used in Sections \ref{sec:level3_2} and \ref{sec:level3_3}, respectivly. In the former, a few nanocrystals can be released using a set-up (Eppendorf Femtojet) which consists of a micropipette (Eppendorf Femtotips) with external diameter of about 2 $\mu$m and inner diameter of 0.5 $\mu$m, held on a 3D micrometeric stage for fine movement. The aqueous suspension of nanocrystals is injected in the micropipette, hence the tip is approached to the region of interest until a micro-drop of suspension is deposited via surface adhesion, with high spatial resolution. In the latter deposition method, a drop of 20 $\mu L$ of the suspension is deposited on the surface followed by a desiccation procedure assisted with vacuum at room temperature, resulting in the deposition of a much larger amount of nanocrystals.

\textbf{Patterned silicon samples preparation}
The suspended shamrock crystal membranes are fabricated on a silicon-on-insulator (SOI) platform with conventional nanofabrication techniques. The SOI chips initially undergo an O$_2$ plasma treatment in an inductively coupled plasma reactive ion etcher (ICP-RIE) at relatively low powers (400 W) for 1 minute to enhance surface adhesion of subsequent resist. CSAR 62 (AR-P 6200.09) positive tone electron beam resist is spun onto the surface at 4000 rpm for 1 minute, and then post-baked at 150$^{\circ}$C for 1 minute. Shamrocks are patterned into the resist with a 30 kV Raith electron beam system and then developed for 1 minute in AR-600-546 developer. The development is stopped with 30 seconds in AR-600-60 stopper, followed by 30 seconds in isopropanol. The pattern is transferred into the device layer with a pseudo-bosch ICP-RIE etch. A low power O$_2$ etch in the ICP-RIE is also able to strip away any residual resist following the silicon etch. Finally, the membrane is suspended in a 3 minute 50$\%$ hydrofluoric acid etch, followed by critical point drying .

\nocite{*}

\bibliography{apssam2}

\end{document}


\preprint{APS/123-QED}

\title{Supplemental Material: \\ Quantum Thermometry with single molecules in portable nanoprobes}

\author{V. Esteso}
 \altaffiliation[Both authors contributed equally]{.}
\affiliation{%
National Institute of Optics (CNR-INO), c/o LENS via Nello Carrara 1, Sesto F.no 50019, Italy}
\affiliation{European Laboratory for Non-Linear Spectroscopy (LENS), Via Nello Carrara 1, Sesto F.no 50019, Italy}
\affiliation{Departamento de Física de la Materia Condensada, ICMSE-CSIC, Universidad de Sevilla, Apdo. 1065, 41080, Sevilla, Spain
}%

\author{R. Duquennoy}
 \altaffiliation[Both authors contributed equally]{.}
\affiliation{%
National Institute of Optics (CNR-INO), c/o LENS via Nello Carrara 1, Sesto F.no 50019, Italy}
\affiliation{%
Physics Department, University of Naples, via Cinthia 21, Fuorigrotta 80126, Italy}

\author{R. C. Ng}%
\affiliation{%
Catalan Institute of Nanoscience and Nanotechnology (ICN2), CSIC and BIST, Campus UAB, Bellaterra, 08193 Barcelona, Spain}

\author{M. Colautti}%
\affiliation{%
National Institute of Optics (CNR-INO), c/o LENS via Nello Carrara 1, Sesto F.no 50019, Italy}
\affiliation{European Laboratory for Non-Linear Spectroscopy (LENS), Via Nello Carrara 1, Sesto F.no 50019, Italy}
 
\author{P. Lombardi}%
\affiliation{%
National Institute of Optics (CNR-INO), c/o LENS via Nello Carrara 1, Sesto F.no 50019, Italy}
\affiliation{European Laboratory for Non-Linear Spectroscopy (LENS), Via Nello Carrara 1, Sesto F.no 50019, Italy}

\author{G. Arregui}%
\affiliation{%
Catalan Institute of Nanoscience and Nanotechnology (ICN2), CSIC and BIST, Campus UAB, Bellaterra, 08193 Barcelona, Spain}
\affiliation{DTU Electro, Department of Electrical and Photonics Engineering, Technical University of Denmark, Ørsteds Plads 343, Kgs. Lyngby, DK-2800, Denmark
}%

\author{E. Chavez-Angel}%
\affiliation{%
Catalan Institute of Nanoscience and Nanotechnology (ICN2), CSIC and BIST, Campus UAB, Bellaterra, 08193 Barcelona, Spain}
 
\author{C. M. Sotomayor Torres}%
\affiliation{%
Catalan Institute of Nanoscience and Nanotechnology (ICN2), CSIC and BIST, Campus UAB, Bellaterra, 08193 Barcelona, Spain}
\affiliation{ICREA, Passeig Lluis Companys 23, 08010 Barcelona, Spain
}%

\author{P. D. Garcia}%
\affiliation{%
Instituto de Ciencia de Materiales de Madrid, (ICMM) Calle Sor Juana Inés de la Cruz 3, 28049 Madrid, Spain. Consejo Superior de Investigaciones Científicas (CSIC)
}%
\author{M. Hilke}%
\affiliation{%
 Department of Physics, McGill University Montréal, QC, Canada H3A 2T8}
\affiliation{Department of Physics, University of Florence, 50019, Sesto Fiorentino, Italy
}%

\author{C. Toninelli}%
\email{Corresponding author: toninelli@lens.unifi.it}
\affiliation{%
National Institute of Optics (CNR-INO), c/o LENS via Nello Carrara 1, Sesto F.no 50019, Italy}
\affiliation{European Laboratory for Non-Linear Spectroscopy (LENS), Via Nello Carrara 1, Sesto F.no 50019, Italy}




\date{\today}




\maketitle

\section{\label{sec:level1}Single-photon emitter molecular thermometer}
The DBT molecules embedded in antracene NCx are well known single-photon emitters widely used for quantum photonics applications. In this experiment, we also perform a measurement of intensity autocorrelation function ($g^{(2)}$) in a Hanbury-Brown Twiss configuration. Figure S1 shows in black the experimental data, whereas the red solid line represents the fitting to a single exponential, leading to $g^{
(2)} (0)$ = 0.02 $\pm$ 0.04 and $\tau_1$ = 3.5 $\pm$ 0.2 ns. The observed antibunching dip shows that we can easily isolate single-quantum emitter within each nanocrystal, with the resonant pumping scheme employed throughout this work.

\begin{figure*}
\includegraphics[width=0.5\columnwidth]{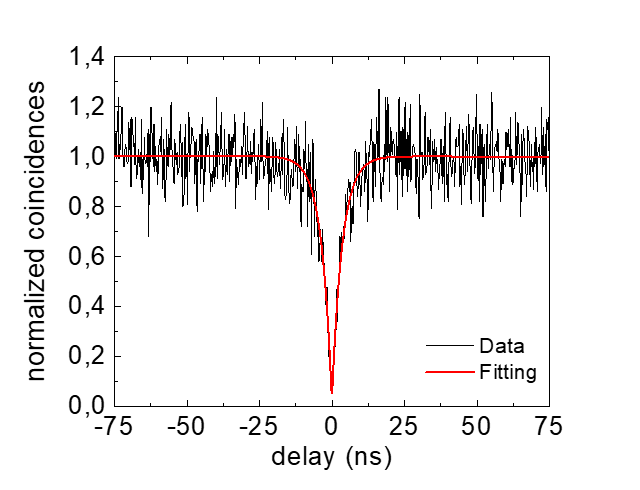}
\caption{\label{fig:S1} Histogram of the relative arrival time between the two SPADs in Hanbury-Brown Twiss configuration. Solid red line is a single exponential
fit to the data, yielding $g^{
(2)} (0)$ = 0.02 $\pm$ 0.04 and $\tau_1$ = 3.5 $\pm$ 0.2 ns.
}
\end{figure*}

\section{\label{sec:level1}Calibration curves of different molecules}
The DBT molecules inside the anthracene NCx can be accomodated in the crystalline host matrix at different positions and relative angles with respect the axis of the lattice structure. This fact, together with the existence of lattice defects, trapped charges, etc..., modify the environment surrounding each DBT molecules and affect the emission of each molecule. Such effect can be noticed in Figure S2 where the calibration curve for three different molecules is displayed, showing that each molecule has a different value of $\Gamma_1/2\pi$. Therefore, the three calibration curves are slightly different but all of them can be fitted (solid lines) with the same $\mu$ and $\xi$ values in Eq. 2 of the main manuscript.

\begin{figure*}
\includegraphics[width=0.5\columnwidth]{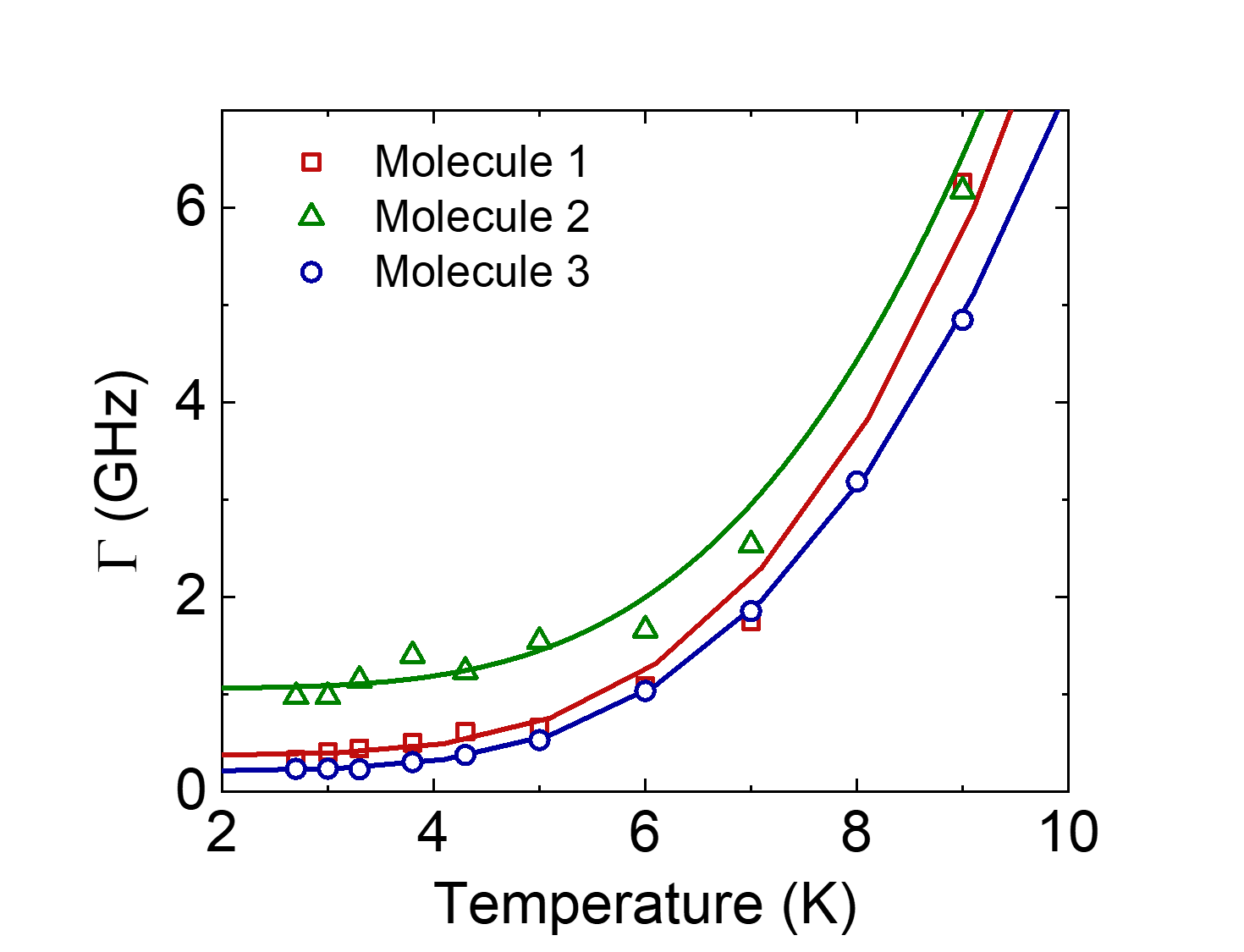}
\caption{\label{fig:S2} Examples of calibration curves of different molecules on a silicon substrate. Scatters stand for the experimental data, whereas solid lines correspond to fitting curves according to Eq. 1 in the main manuscript. The fixed values of parameters $\mu$ and $\xi$ are (7.1 $\pm$ 0.5) x 10$^{-6}$ ps$^5$ and (6.1 $\pm$ 0.3) ps$^{-1}$, respectively, which are the same values as in the main manuscript.
}\end{figure*}


